\documentclass[a4paper,11pt]{article}
\usepackage{pos_1}
\usepackage{bm}
\usepackage{amsmath}

\title{Oscillations of Majorana neutrinos in supernova and CP violation}

\author*[a]{Artem Popov}
\author[a]{Alexander Studenikin}

\affiliation[a]{Department of Theoretical Physics, \\ Moscow State University, 119991 Moscow, Russia}

\emailAdd{ar.popov@physics.msu.ru}
\emailAdd{studenik@srd.sinp.msu.ru}

\abstract{Leptonic CP violation is one of the most important topics in neutrino physics. CP violation in the neutrino sector is strongly related to the nature of neutrinos: whether they are Dirac or Majorana particles. In \cite{Popov:2021icg} we have shown that for Majorana neutrinos arance of nonzero Majorana CP-violating phases combined with strong magnetic field during supernova core-collaple can induce new resonances in neutrino oscillations, for example in $\nu_e \to \bar{\nu}_\tau$ channel. In this contribution we further study new resonances in Majorana neutrino oscillations, in particular energy dependence of amplitudes of these resonances. Our findings suggest a potential astrophysical setup for studying the nature of neutrino masses and leptonic CP violation and may be important for future neutrino experiments, such as JUNO, Hyper-Kamiokande and DUNE.}

\FullConference{%
  41st International Conference on High Energy physics - ICHEP2022\\
  6-13 July, 2022\\
  Bologna, Italy
}

\begin{document}
\maketitle
\section{Majorana neutrino oscillations in supernova media}
It is well known that mixing matrix for the case of Majorana neutrinos can be written as follows
\begin{equation}
	U = U_{PMNS}(\theta_{12}, \theta_{13}, \theta_{23}, \delta) \cdot \text{diag}(e^{i\alpha_1}, e^{i\alpha_2}, 1),
\end{equation}
where $ U_{PMNS}$ is the Pontecorvo–Maki–Nakagawa–Sakata matrix, $\theta_{ik}$ are mixing angles, $\delta$ is Dirac CP-violating phase, and $\alpha_1$ and $\alpha_2$ are Majorana CP-violating phases that only present in case if neutrinos are Majorana particles. In our paper \cite{Popov:2021icg} we have studied the process of Majorana neutrino oscillations in environment peculiar for supernova explosion, in particular strong magnetic field ($10^{12}$ Gauss and more) and dense matter. It was shown that presence of nonzero Majorana CP-violating phases can modify patterns of neutrino-antineutrino oscillations induced by the supernova magnetic field.

To study neutrino oscillations in supernova environment we numerically solve the following equation
\begin{equation}\label{dirac_equation}
(i\gamma^{\mu} \partial_{\mu} - m_i - V^{(m)}_{ii} \gamma^{0}\gamma_5)\nu_i(x) -\sum_{k \neq i} \big(\mu_{ik}^M \bm{\Sigma}\bm{B} + V^{(m)}_{ik}  \gamma^{0}\gamma_5\big)\nu_k(x) = 0,
\end{equation}
where 
\begin{equation}
	V^{(f)} = \operatorname{diag} \left(\frac{G_F n_e}{\sqrt{2}} - \frac{G_F n_n}{2\sqrt{2}}, - \frac{G_F n_n}{2\sqrt{2}}, - \frac{G_F n_n}{2\sqrt{2}} \right)
\end{equation}
is the Wolfenstein potential, $V^{(m)} = U^{\dag}V^{(f)}U$ is the matter potential in the massive neutrinos basis, and $n_e$, $n_n$ are electron and neutron number densities respectively. The magnetic moments matrix of Majorana neutrinos $\mu^M$ from Eq. (\ref{dirac_equation}) is antisymmetric and imaginary \cite{Giunti:2014ixa}, and can be parametrized in the following form
\begin{equation}
	\mu^M = \begin{pmatrix}
		0 && i\mu_{12} && i\mu_{13} \\
		-i\mu_{12} && 0 && i\mu_{23} \\
		-i\mu_{13} && -i\mu_{23} && 0
	\end{pmatrix},
\end{equation}
where $\mu_{ij}$ are real transition magnetic moments.
The values of neutrino magnetic moments are presently unknown. In \cite{Canas:2015yoa}, using solar neutrino data from Borexino, the authors found the following upper bounds for transition magnetic moments Majorana neutrino
\begin{eqnarray}
	\mu_{12} &=& 3.1 \times 10^{-11} \mu_B, \nonumber \\
	\mu_{13} &=& 4.0 \times 10^{-11} \mu_B, \;\;\;\; \text{(90\% C.L.)}\\
	\mu_{23} &=& 5.6 \times 10^{-11} \mu_B. \nonumber
\end{eqnarray}

\begin{figure}[h]
	\begin{minipage}[h]{0.49\linewidth}
		\center{\includegraphics[width=1\linewidth]{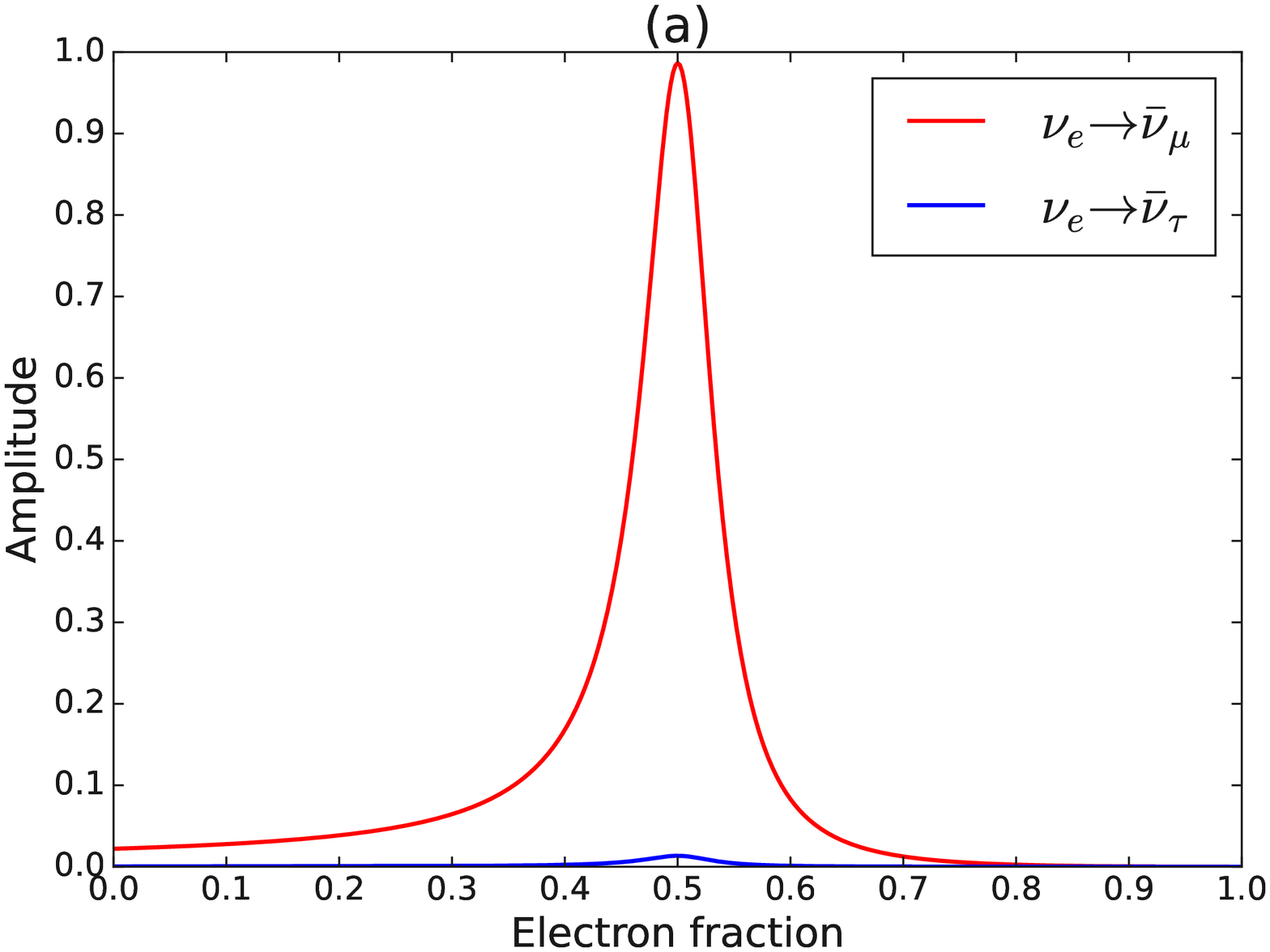}}
	\end{minipage}
	\begin{minipage}[h]{0.49\linewidth}
		\center{\includegraphics[width=1\linewidth]{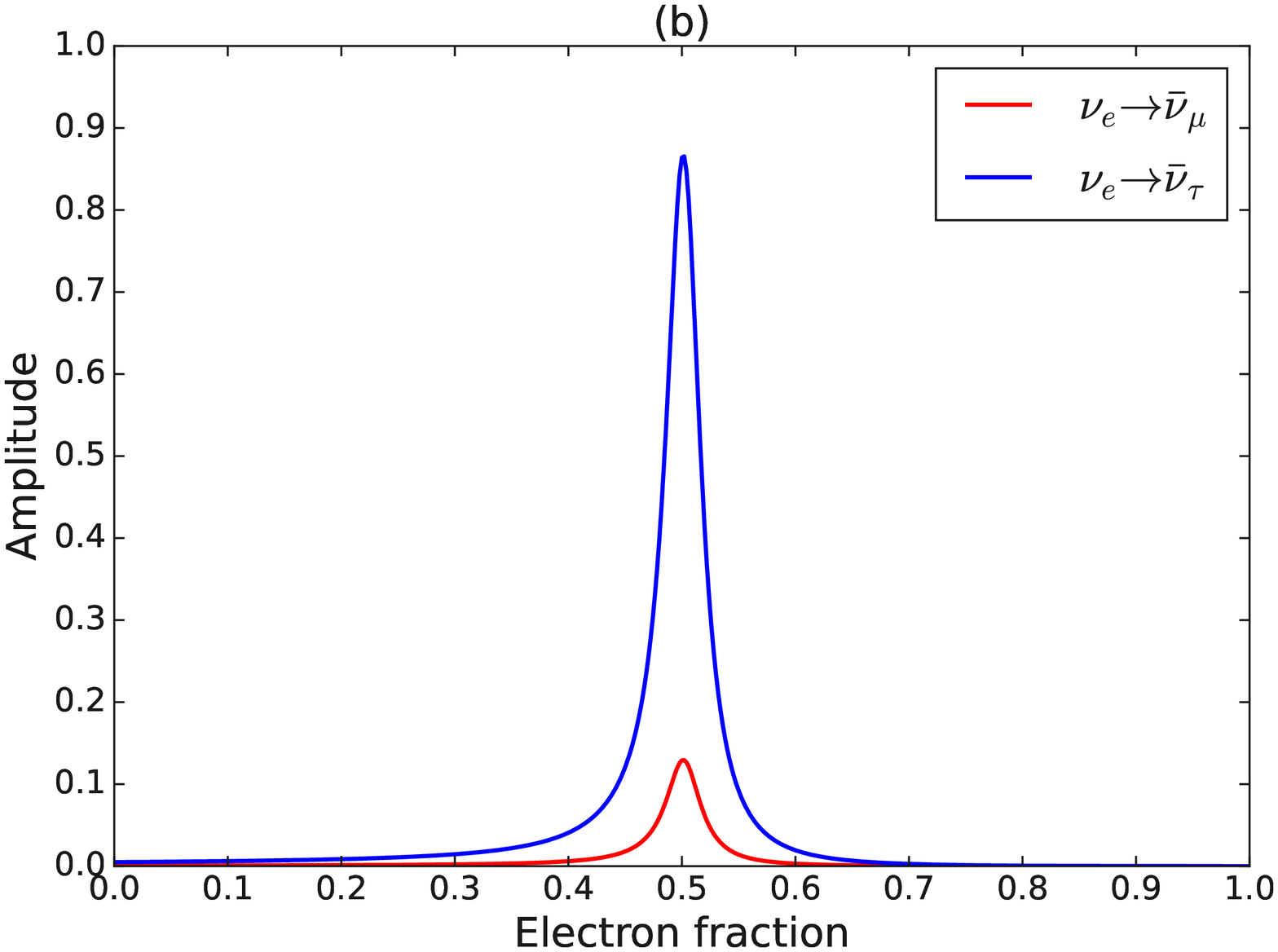}}
	\end{minipage}
	\vspace{-3mm}
	\caption{Amplitudes of Majorana neutrino oscillations in supernova media for $B = 10^{12}$ Gauss as functions of electron fraction $Y_e$. (a) $\alpha_1 = 0$, $\alpha_2 = 0$; (b) $\alpha_1 = \pi$, $\alpha_2 = \pi$.}
	\label{fig1}
	\vspace{-5mm}
\end{figure}

Using Eq. (\ref{dirac_equation}), we numerically study the amplitudes of neutrino-antineutrino oscillations. Here we take realistic values for supernova magnetic field $B = 10^{12}$ Gauss and matter density $n_n = 10^{30}$ cm$^{-3}$. For simplicity we also assume $\mu_{12} = \mu_{13} = \mu_{23} = 10^{-12} \mu_B$. Figure \ref{fig1}a shows the amplitudes of $\nu_e \to \bar{\nu}_\mu$ and $\nu_e \to \bar{\nu}_\tau$ oscillations as functions of electron fraction $Y_e = n_e/(n_n + n_e)$ for the case $\alpha_1 = 0$ and $\alpha_2 = 0$. The amplitude of $\nu_e \to \bar{\nu}_\tau$ is close to zero, while $\nu_e \to \bar{\nu}_\mu$ oscillations undergo resonant enhancement at $Y_e = 0.5$. This well known phenomenon of resonant spin-flavour conversion was proposed for the first time in \cite{Akhmedov:1988uk,Lim:1987tk}. In Figure \ref{fig1}b we show the amplitudes of neutrino-antineutrino oscillations for the case of nonzero Majorana CP-violating phases, namely $\alpha_1 = \pi$ and $\alpha_2 = \pi$. In this case resonant enhancement in $\nu_e \to \bar{\nu}_\tau$ oscillations appears instead of $\nu_e \to \bar{\nu}_\mu$ oscillations at $Y_e = 0.5$. In \cite{Popov:2021icg} we have shown that this new resonance can result in potentially observable phenomena during supernova core-collapse, in particular to modification of $\nu_e$ and $\bar{\nu}_e$ fluxes ratio.

\begin{figure}[h]
	\begin{minipage}[h]{0.49\linewidth}
		\center{\includegraphics[width=1\linewidth]{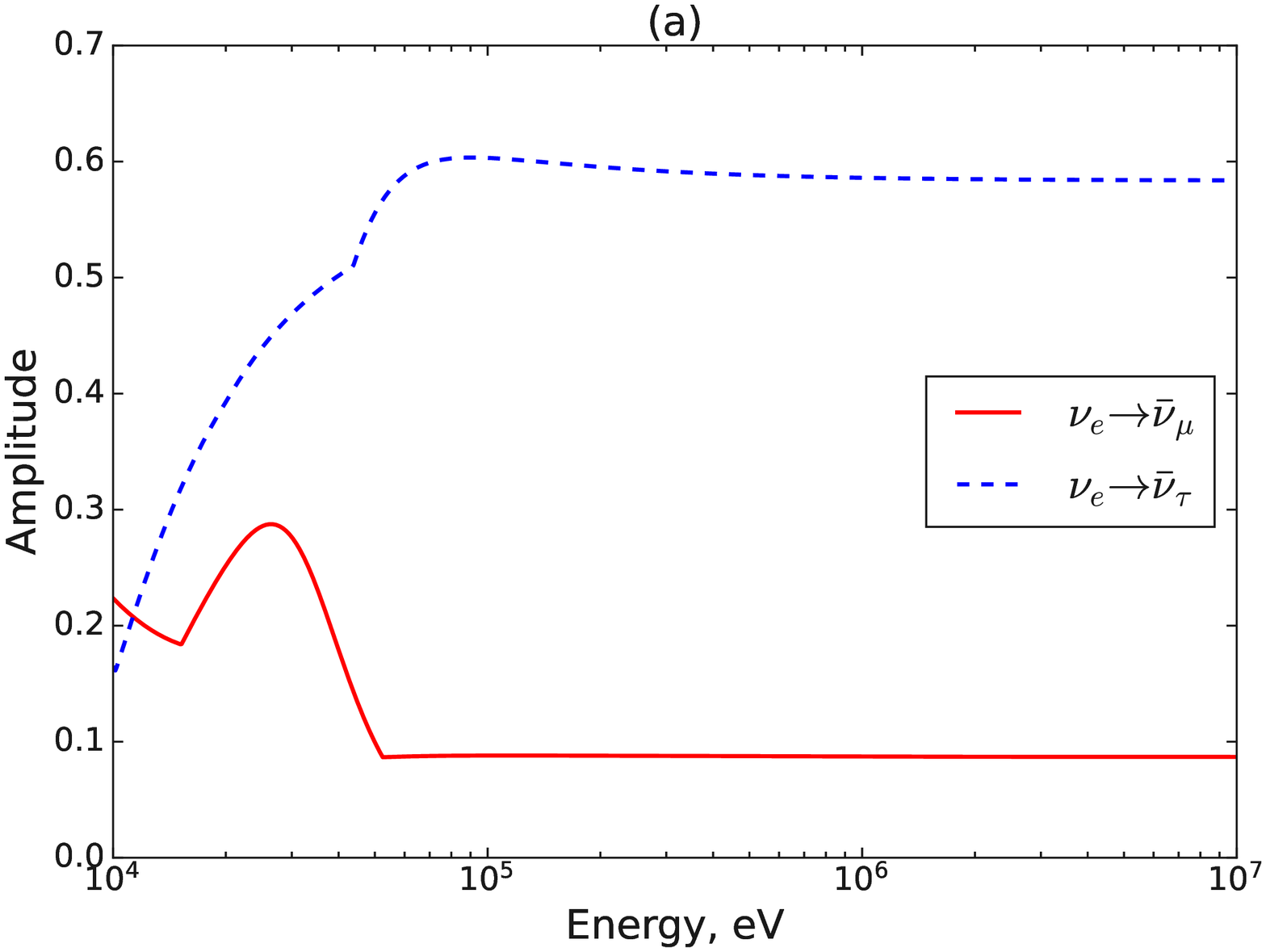}}
	\end{minipage}
	\begin{minipage}[h]{0.49\linewidth}
		\center{\includegraphics[width=1\linewidth]{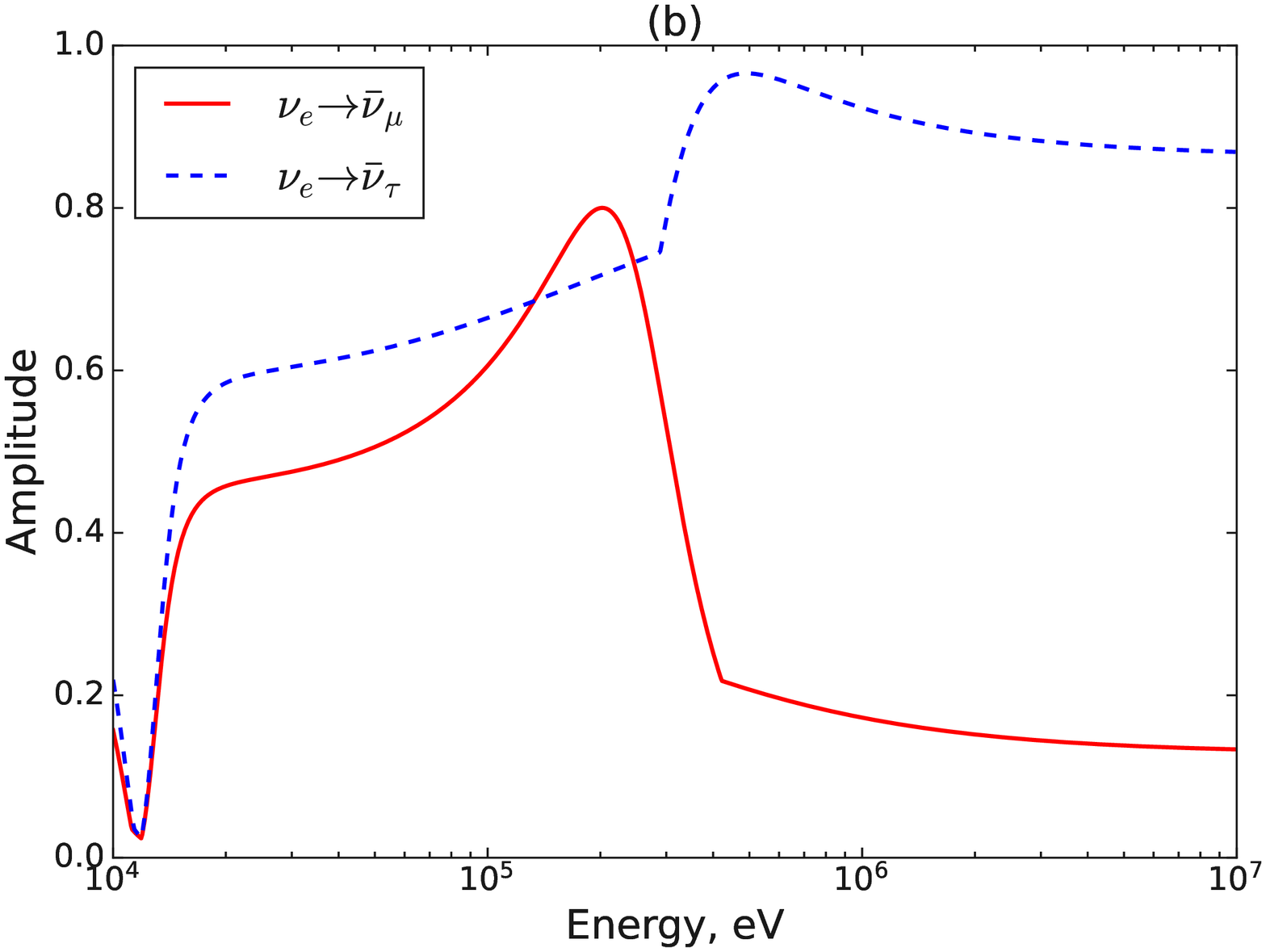}}
	\end{minipage}
	\vspace{-3mm}
	\caption{Amplitudes of resonant conversion $\nu_e \to \bar{\nu}_\tau$ at $Y_e = 0.5$ for $B = 10^{12}$ Gauss as functions of neutrino energy. (a) $\mu_\nu = 10^{-11} \mu_B$; (b) $\mu_\nu = 10^{-12} \mu_B$.}
	\label{fig2}
	\vspace{-5mm}
\end{figure}

Throughout \cite{Popov:2021icg} we assumed that neutrino energy is constant $E = 10$ MeV, which is considered to be a typical value for supernova neutrinos. It is also important to study behaviour of new resonance at different neutrino energies. Figure \ref{fig2}a and Figure \ref{fig2}b show the amplitudes of resonant $\nu_e \to \bar{\nu}_\tau$ oscillations as functions of neutrino energy $E$ for magnetic moment values of $\mu_\nu = 10^{-11}\mu_B$ and $\mu_\nu = 10^{-12}\mu_B$ respectively. We notice that in both cases resonant enhancement is indeed observed for supernova neutrinos with energy of 1 MeV and higher.

\section{Conclusion}
In this contribution we have studied the process of Majorana neutrino oscillations in supernova media, in particular in a strong magnetic field. We have shown that for certain nonzero values of Majorana CP-violating phases $\alpha_1$ and $\alpha_2$ resonant enhancement of $\nu_e \to \bar{\nu}_\tau$ oscillations can occur in region of supernova media where electron fraction $Y_e = 0.5$. Note that $Y_e = 0.5$ is expected to be realized during supernova core-collapse (see \cite{Buras:2005rp}). We show that this resonant amplification is present for realistic supenova neutrino energies, but it can disappear for energies lower that $\sim$ 100 keV.

\section*{Acknowledgements}
The work is supported by the Russian Science Foundation under grant No.22-22-00384. The
work of A.P. has been supported by the Foundation for the Advancement of Theoretical Physics and
Mathematics “BASIS” under Grant No. 21-2-2-26-1 and by the National Center for Physics and
Mathematics (Project “Study of coherent elastic neutrino-atom and -nucleus scattering and
neutrino electromagnetic properties using a high-intensity tritium neutrino source”).

\end{document}